\documentclass[convention,peer-reviewed]{aesconf}

\graphicspath{{./}{figures/}}

\usepackage[utf8]{inputenc}

\usepackage{microtype}

\usepackage[numbers,square]{natbib}

\usepackage{booktabs}
\usepackage{multirow}
\usepackage{color}
\usepackage{url}
\usepackage{tabularx}

\usepackage{hyperref}
\usepackage{cleveref}
\usepackage[hypcap=true]{caption}

\title{Adoption of AI Technology in the Music Mixing Workflow: An Investigation}

\author[1]{Soumya Sai Vanka}
\author[2]{Maryam Safi}
\author[2]{Jean-Baptiste Rolland}
\author[1]{Gy\"{o}rgy Fazekas}

\affil[1]{Queen Mary University of London, London, UK}
\affil[2]{Steinberg Media Technologies GmbH, Hamburg, Germany}

\correspondence{Soumya Sai Vanka}{s.s.vanka@qmul.ac.uk}

\lastnames{Vanka et al.}

\shorttitle{Investigation of AI tools in mixing workflow}

\savebox{\AEStop}{%
	\begin{minipage}{\textwidth}%
		\rule{\textwidth}{1.5pt}\\%
		\\%
		\begin{minipage}[c][\iftoggle{convention}{3.2cm}{3.7cm}][t]{0\textwidth}%
			\includegraphics[width=20mm]{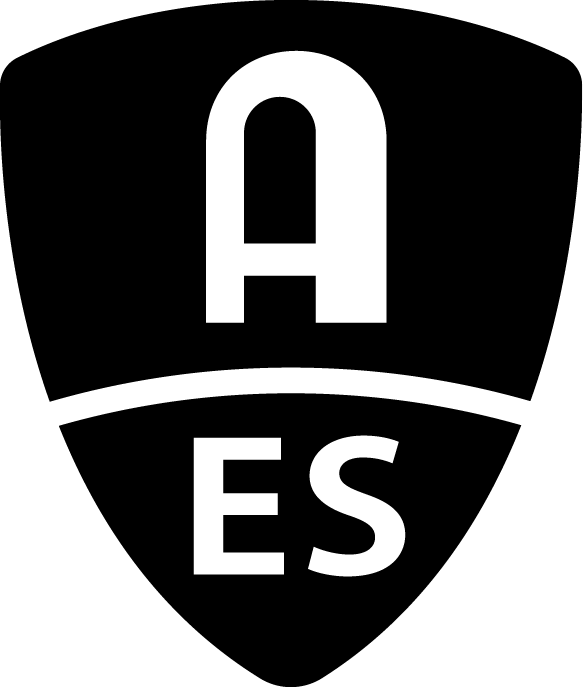}%
		\end{minipage}%
		\begin{minipage}{\textwidth}%
			\sffamily%
			\begin{center}%
				\LARGE Audio Engineering Society\\%
				\iftoggle{express_paper}{%
				\hspace{3mm}\fontsize{36}{38pt}\selectfont Convention Express\\Paper \AESExpressPaperNumber\\%
				}{%
				\iftoggle{convention}{%
				\fontsize{36}{38pt}\selectfont Convention Paper\\%
				}{%
				\fontsize{36}{38pt}\selectfont Conference Paper\\%
				}}%
				\vspace{0.2cm}%
				\large Presented at the \AESConferenceNumber \iftoggle{convention}{Convention\\}{\AESConferencePrefix Conference on\\}%
				\iftoggle{convention}{}{\AESConferenceTopic\\}%
				\AESConferenceDate\ifx\AESConferenceLocation\empty\else, \AESConferenceLocation\fi%
			\end{center}%
		\end{minipage}\\%
		\vspace{0.2cm}\\%
		\begin{minipage}{\textwidth}%
			\rmfamily\itshape\small	\AESLegalTextPrefix\ \AESLegalText%
		\end{minipage}\\%
		\\%
		\rule{\textwidth}{1.5pt}%
	\end{minipage}%
}

\begin{document}

\twocolumn[
\maketitle 

\begin{onecolabstract}
The integration of artificial intelligence (AI) technology in the music industry is driving a significant change in the way music is being composed, produced and mixed. This study investigates the current state of AI in the mixing workflows and its adoption by different user groups. Through semi-structured interviews, a questionnaire-based study, and analyzing web forums, the study confirms three user groups comprising amateurs, pro-ams, and professionals. Our findings show that while AI mixing tools can simplify the process and provide decent results for amateurs, pro-ams seek precise control and customization options, while professionals desire control and customization options in addition to assistive and collaborative technologies. The study provides strategies for designing effective AI mixing tools for different user groups and outlines future directions.
\end{onecolabstract}
]

\section{Introduction}
The field of AI-assisted music production has seen a significant increase in recent years, with a growing number of tools and techniques aimed at automating various aspects of the process. However, despite the potential benefits of these tools, many music producers and engineers remain skeptical of their ability to match the performance of human experts. This skepticism is driven by several factors, including fear of being replaced~\citep{AINow, sturm2019artificial}, doubt that machines can perform as well as humans on subjective tasks~\citep{jordanous2019evaluating, ritchie2019evaluation, lake2017building}, and distrust of AI recommendations due to a lack of understanding of the underlying reasoning behind the models' outputs~\citep{ribera2019can, roy2019automation}. These reasons pose a major bottleneck to the adoption of these tools in users' workflows.

In this investigation, we examine the attitudes and perceptions of various user groups who employ AI technologies in mixing workflows and describe some use case scenarios and expectations from AI tools for each user type. This is achieved by conducting thorough interviews with professional engineers (pros), followed by a questionnaire with pros and professional-amateurs (pro-ams). We also qualitatively assess the discussions on various internet forums to construct the bigger picture about the sentiments around these tools amongst a wider segment of users including beginners and amateurs. We are confident that the findings of this study will motivate the creation of suitable AI-based tools for mixing workflows that specifically address the needs and specifications of each user group in accordance with their expectations. 

\section{Background}
\subsection{Music Mixing}
Mixing music is the process of bringing multiple audio tracks together to form a cohesive final product that evokes emotions and tells a story~\citep{miller2016mixing}. The job of the mixing engineer is to take the raw audio recordings and perform technical and creative transformations to align them with the artist's vision and create a polished, cohesive final mix. This includes tasks such as highlighting key elements, balancing instrumentation, and creating a unique sound. In addition to gain staging, techniques such as equalization, panning, compression, and modulation effects are used to shape the sound~\citep{izhaki2017mixing}. The mixed music is then passed on to the mastering engineer for further enhancements before it is ready for commercial release.

\subsection{Democratisation of Music Production}
\label{sec:user}
Before the advent of digital technologies, recording and producing music required expensive equipment and specialised skills that were beyond the reach of most musicians. This meant that those with financial resources and infrastructural support had a better chance to access professional studios and distribute their music. However, with the introduction of personal computers, the internet and digital audio workstations (DAWs), the cost of recording, producing, and distributing music has significantly decreased, making it accessible to a much wider range of users including musicians~\citep{hracs2012creative}. The removal of the traditional barriers of cost and skill has allowed a larger number of musicians/artists/producers to enter the industry, creating a more diverse segment of user groups for these tools. The following section introduces the three categories of users identified from existing literature~\citep{sandler2019semantic, mcgrath2016making, bromham2016can}. 

\subsubsection{Amateurs}
An amateur in the context of mix engineering is someone who is new to the field and may not have formal training or experience in mixing. They may be hobbyists or musicians who are just starting out and learning the basics of mixing music~\citep{hoare2014coming}. They are typically less skilled and do not get paid for mixing music. Users may continue to remain in this category for a prolonged period of time if they only use their mixing skills occasionally.  

\subsubsection{Pro-Ams}
Robert Stebbins, a sociologist coined the term ``modern amateur'' as someone who is skilled, dedicated and passionate but different from professionals and hobbyists~\citep{stebbins1977amateur}. More recent discussions on amateurism have recognized the impact of societal changes and the internet on the concept of amateurism, and have introduced the term ``Pro-Am'' to describe individuals who are seriously engaged in a field and work to professional standards but without the same infrastructural support enjoyed by professionals~\citep{prior2018new}. The Pro-Am is seen as a spectrum between full-fledged professionals and traditional amateurs~\citep{leadbeater2004pro}. They may have a passion for making and mixing music and a desire to improve their skills, but they may not be making a living from mixing music directly or this may not be their main focus in the field. 

\subsubsection{Professionals}
A professional mixing engineer is an individual who is highly skilled in mixing music and is paid for their services. They have possibly received formal education or have tacit knowledge gained from a formal apprenticeship or internship in a traditional studio environment~\citep{bromham2016can}. They have good connections in the music industry and have the infrastructural support to afford equipment and facilities. They are also expected to continuously improve and update their knowledge and skills to stay current in their field. They are considered experts in mixing and are held to a higher standard of performance than amateurs or hobbyists. 

\subsection{Smart tools for mixing workflows}
In 1975, Dan Dugan presented an automatic mixing system at an Audio Engineering Society convention that focused on microphone mixing to reduce feedback~\citep{dugan1975automatic}. Ever since, there has been extensive research in building assistive mixing tools and automatic mixing systems using knowledge-based, psychoacoustics, and perceptually-motivated approaches~\citep{tenyearsai, steinmetz2022automix}. With the success of machine learning and artificial intelligence (AI) in computer vision and other fields, these techniques have been applied to the audio domain including building smart tools for music mixing. These tools have ranged from mixing systems that take in stems and provide a mix~\citep{tom2019automatic, moffat2019machine,martinez2021deep,steinmetz2021automatic, colonel2021reverse, de2013knowledge, steinmetz2020learning, martinez2022automatic, koo2022music} to AI-based plugins that emulate their analogue/digital counterparts and apply processing directly to provided audio by analysing them~\citep{ramirez2018end,sheng2019feature, singh2021intelligent, ma2015intelligent, steinmetz2022efficient, kuznetsov2020differentiable, colonel2022direct, nercessian2020neural, colonel2022reverse, lee2022differentiable}. Plugins that employ rule-based processing have also been proposed previously~\citep{Wilmering:2012fk}.

Smart or AI-based tools aim to provide solutions for various tasks in the mixing workflows using state-of-the-art statistical software technology from the field of artificial intelligence. The development of smart plugins and tools for music mixing is motivated by the goal of automating technical and non-creative tasks in the mixing process, thus reducing the time and skills needed to mix a song~\citep{tsiros2020towards}. Additionally, such tools aim to provide convenience and efficiency in the mixing process, allowing users to spend more time on the creative aspects of music production~\citep{tenyearsai, de2019intelligent}.

 Several AI-based commercial tools for music-mixing workflows have been made available in the market in recent years. For the convenience of discussion, we divide smart tools available for mixing workflow into two categories.
 \begin{itemize}
    \item \textbf{Automatic mixing}: Automatic mixing can be defined as any system that can create a mix given the stems or raw audio. In other words, Automatic mixing is a broader term that encompasses the use of AI to automate the entire mixing process. RoEx\footnote[2]{\url{https://www.roexaudio.com/}} and iZotope Neutrone\footnotemark[1] offer automatic mixing services and tools, respectively.  
    \item \textbf{Assistive mixing tools}: Assistive mixing tools use AI technology to assist in mixing workflows. These tools include AI-based plugins and processors such as smart equalizers, smart reverbs, smart vocal riders, etc. Such tools are offered by several plugin companies including iZotope\footnotemark[1], Sonible\footnote[3]{\url{https://www.sonible.com/}}, Focrusite\footnote[4]{\url{https://focusrite.com/en}}, and more.
\end{itemize}

\label{factors}
However, the success of these tools in the marketplace depends on three factors as well explained in ~\citep{tsiros2020towards}:
\begin{enumerate}
    \item \textbf{Precision and quality of the produced results}: Audio generated by AI-based models needs to be of high quality in order to be used by professionals in their workflows. However, current AI-based models still often struggle to generate high-quality output which causes resistance to their adoption in professional workflows.
    \item \textbf{Seamless integration into existing workflows}: Professionals tend to develop a workflow in their practice over years of iteration that is best suited to maximise their efficiency and productivity. For the better acceptance of these tools, seamless integration into existing individual workflow is key~\citep{long2021role}.
    \item \textbf{Interaction models that facilitate trust}: The biggest inhibition towards the use of AI-based tools is rooted in them being perceived as ``black boxes'' that lack interpretability~\citep{linardatos2020explainable, gilpin2018explaining}. The lack of control for alteration to achieve desirable results is often a caveat for many professional engineers~\citep{sterne2019machine, scurto2018appropriating}. 
\end{enumerate}

In this work, we aim to understand the adoption of AI-based tools in music mixing workflows by identifying user groups and analyzing their use cases, expectations, and sentiments. Utilising semi-structured interviews, a questionnaire-based study, and internet forum analysis, we gather data and perform a qualitative analysis. The findings are presented and analysed, followed by a discussion on potential future directions for the design of AI-based mixing tools. The research concludes with a summary of the findings and relevant future directions.


\section{Methodology}
To counter for the diverse views and insights from individuals with varying skill levels in mixing, we deployed three mechanisms for data collection, namely structured, semi-structured, and unstructured, not in this particular order~\citep{seale2010interviews}. The benefits of each of these methods have been described in the literature~\citep{brinkmann2014unstructured} in detail. Each of the methods are somewhat complementary in terms of information derived from them. 

\subsection{Semi-Structured Interviews with Pros}
Expert knowledge can be used for modelling the world~\citep{shaw1990modeling}. While this viewpoint may be simplistic from an interpretive-constructive perspective, it is useful for conducting interviews to collect information. To start, we conducted semi-structured interviews with mixing engineers who are professionals in their field. To avoid the problem of selectivity which is very important for the validity of the investigation, multiple interviews (five) were conducted~\citep{bogner}. Since we targeted experts at this stage of the study, who represent a significantly smaller number of users than non-experts, and who are difficult to reach, five expert participants were personally identified amongst our contacts based on the criteria described in \Cref{sec:user} and invited to participate in individual interviews~\citep{baker2017many}. All expert engineers have been practising mixing for more than five years, and are considered advanced in their skills. They mix music for themselves and other artists. These mixing engineers have received formal training, get paid for their work, and have had their work published as traditional discographies and on the music streaming platforms . The objective for conducting interviews was mainly to facilitate some preliminary theory-building. The interviews focused on understanding the use, expectations, and sentiments around AI-based technology in mixing workflows amongst the participant engineers. The interviews were conducted via online video calls. These calls were recorded and then transcribed. We performed thematic analysis on the transcripts to develop themes as elaborated in \Cref{sec:data_analysis} 

\subsection{Questionnaire-based Structured Study with Pros and Pro-ams}
A structured and standardised way of asking questions leads to comparable answers across participants~\citep{brinkmann2014unstructured}. We aimed to derive quantifiable data based on the themes developed from the interviews and check for agreement amongst a larger pool of mixing engineers. Hence, we conducted a questionnaire-based structured study with invited pro and pro-am mixing engineers based on the criteria described in \Cref{sec:user}. This study involved twenty-two participant engineers who considered their skill level in mixing to be advanced to expert and had more than five years of experience. The questionnaire comprised multiple choice questions and free form long answer questions.

\subsubsection{Data from Internet Forums on AI-based Mixing and Mastering}
Further, to expand the horizons of our results, we scraped textual data from discussions on various internet forums such as Reddit, Quora, and Twitter that were specifically focused on AI-based tools for mixing and mastering workflows~\citep{bickart2001internet, seale2010interviews}. We collected data from nearly six different discussions, involving about thirty participants across all the threads. Since we were looking for recent discussions in this niche field, we could only find a limited number of threads. Also, the participant profiles were not clear in many of these discussions. We assumed participants to be either amateurs or pro-ams. 

\subsection{Data Analysis}
\label{sec:data_analysis}
The collected data from different studies was both qualitative and quantitative in nature. Hence, the data were evaluated based on a mixed method, using both qualitative and quantitative research approaches.
The semi-structured interviews were recorded and transcribed. We performed qualitative thematic analysis on the transcripts using grounded theory analyses~\citep{braun2012thematic, braun2006using} that involved coding of the data, customisation of the code system, and construction of theories iteratively and inductively. We also performed thematic analysis on all the qualitative data collected from the questionnaire and forums using similar inductive, iterative, and grounded approaches. Qualitative analyses were completed in the MAXQDA\footnote[5]{\url{https://www.maxqda.com/}} software. MAXQDA is designed for computer-assisted qualitative and mixed methods data, text and multimedia analysis in academic, scientific, and business institutions.
We first generated a series of ``codes'' that identified different subgroups of users of AI mixing tools. Over several iterations of the process, we deduced their use-case and sentiments around these tools. The multiple-choice questions from the questionnaire were visualised using numerical and statistical methods which in many cases complemented and strengthened the qualitative findings. The outcomes will be discussed in the following sections. 


\section{Results}

\begin{figure}
    \centering
    \includegraphics{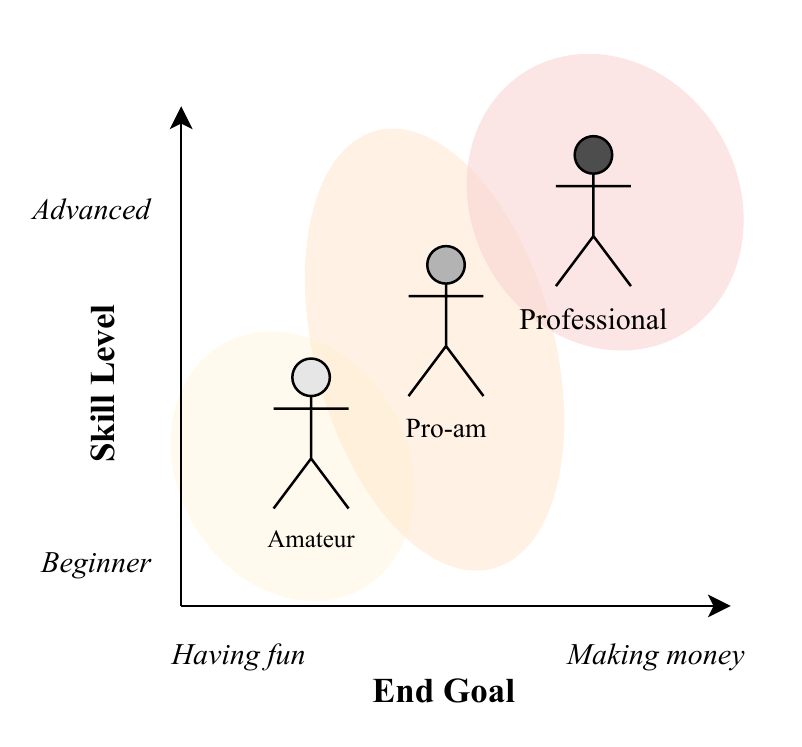}
    \caption{Categories of users for smart tools}
    \label{fig:users}
\end{figure}

Our investigation confirms the relevance of AI-based mixing tools in all three distinct groups of users: amateurs, pro-ams, and professionals as shown in \Cref{fig:users}. We found that each user group has unique requirements for the intelligent tools and employs them for various purposes in their workflows. We discuss the results of all three studies together. We quote excerpts from semi-structured interviews and questionnaires indicated by 'I' and those from internet forums by 'F'. 

\subsection{Amateurs}
We observed that many forum participants who identified as amateurs were primarily musicians who create and compose music. They indicated limited knowledge of music mixing and their main objective is to share their music with the world.

\textit{``The biggest barrier to me getting a recording out into the world is always the mixing process. It's tedious, ... Give me a button I can push to automate most of the work.''}

Several amateurs noted that mixing is the biggest hurdle in putting their music out. They also acknowledged that they are happy if the mix is decent enough.

\textit{``...give a beginner a simply decent mix. I'm not expecting to sound 100\% perfect, but if they help, it would be enough for me.''}

Fully Automatic mixing tools are more useful to this group since they typically don't have the time or resources to invest in expensive equipment or hire a professional engineer. These tools can be a cost-effective solution for hobbyist musicians whose primary goal is creating and sharing music.

\textit{``Getting a mix or master 90\% listenable is good enough for me to release music and lets me concentrate on what I enjoy.''} [F]

Amateurs in the music production industry see AI-powered tools as a way to save time and trouble, and to produce a decent mix with minimal effort without a deep understanding of the technical aspects of mixing. Due to their limited knowledge of the subject, they expect these systems to be highly autonomous almost like a ``one-push magic button''. They also have lower expectations of the quality of the mix produced by these AI systems.

\textit{``I love recording my own songs, but I really can't understand many of the concepts regarding mixing, I get easily lost with EQs, compression, etc. I'm basically an amateur, I respect the art of mixing, but that's the phase where I lose the fun of making music.''}  [F]

Overall, we observed that amateurs are positively embracing this emerging technology.

\subsection{Pro-ams}
Our observations confirmed that pro-ams or semi-professional users have a higher level of technical skill than amateurs, but may still not have the same level of experience and expertise as professionals. Also, most often they don't make money out of the craft, rather they produce and mix music for joy. They use smart mixing tools in a similar way to amateurs, but they also use them as a way to improve their skills and work towards becoming professional mix engineers. Pro-ams also use smart mixing tools as a way to quickly achieve a certain sound or style in their mixes. 


\textit{``It could be used as a tool to learn the basics, then using the reference method, as well as trial and error, it'll be another tool in the tool belt of learning. AI being used for basics, other methods to make more exciting mixes.''} [F]

Given their higher level of expertise, pro-ams are well aware of the limitations of this technology but are willing to maximise the available smart tools to the best of their potential. They are cautiously optimistic about the future of these tools.

\textit{``None of this AI mixing is going to mix your music for you. It’s going to make basic suggestions of moves you can make to better process the audio. You, yourself, still have to take the brunt of the decision-making and the work.''} [F]

\begin{figure}
    \centering
    \includegraphics[scale=0.8]{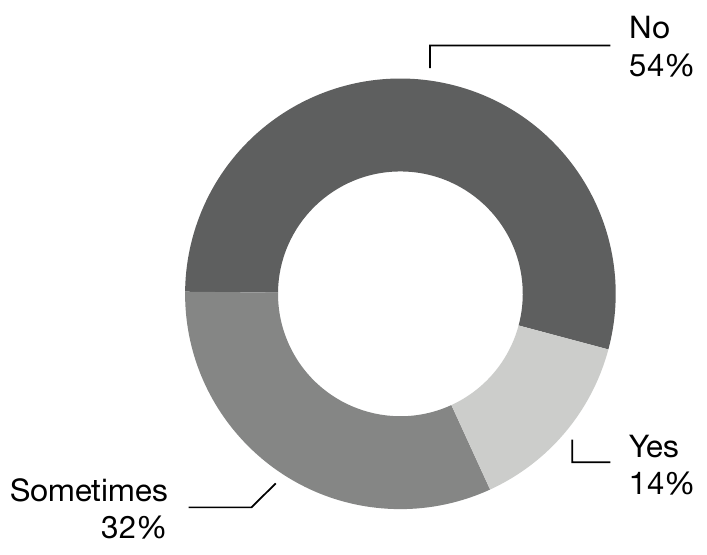}
    \caption{Responses to the use of AI-powered tools in mixing workflow as reported by pro-ams and pros.}
    \label{fig:usage}
\end{figure}

\subsection{Professionals}
When it comes to the experts in the field, professional engineers have a mixed view on the use of AI in mixing workflows. Some professionals use smart mixing tools as a way to improve their workflow and save time on repetitive tasks. 

\textit{``How can we best use our time when we're mixing a record or when we're producing a record and having intelligent tools that take away some of these mundane tasks?''} [I]

They may use these tools to assist them in the mixing process but ultimately rely on their own expertise and experience to create a polished final product. They believe these tools can provide a starting point and direction for the mix. Professionals may also use these tools as a way to experiment with new sounds and styles, or as a way to quickly achieve a certain sound or style in their mixes.

\textit{``You maybe can think of five amazing creative ideas in the mix if you're lucky, whereas the system can maybe hit you with 100 great ideas and you can choose''} [I]

 \textit{``I've used Neutron's masking feature to get some extra clarity but your talking one move of the potentially hundreds of thousands of moves involved in a production.''} [I]

They expect these tools to be highly accurate, precise and customisable to their needs. They are certainly in favour of assistive and co-creative technologies that enable collaboration where the machine makes suggestions and the engineer can then adjust and fine-tune settings to their liking.


 \textit{``...you know comparing your mix to what is going on on the reference song and recommending there is too much low end, there's too much higher end, there's too much of this and that.''} [I]

\textit{``Machine could process as well and then give me the controls to just adjust everything. It would be much more helpful than starting from scratch.''} [I]

Our investigation revealed that pros are currently using AI-based smart tools for tasks such as filtering, peak detection, pitch detection, mastering, equalisation, and sound enhancement. 




However, a large number of professional engineers from our study said that they do not use smart AI tools in their mixing workflows as shown in \Cref{fig:usage}. These experts argue that AI-powered tools can't fully replace the human touch and creativity required in the mixing process. They argue that these tools generate outputs using computational methods and do not understand the `feelings` and `emotions` required for mixing music intentionally. They argue that AI-controlled mixing could lead to a loss of control and precision in the final product. 

\textit{``For a hobby, they can do that at home. On a professional level in a production setting, it won’t fly because client and corporate revisions are demanding and very precise in what they want on a granular level. It’s NOT about doing it faster but with precision.''} [F]

Another expert pointed out that these tools are often met with scepticism in the court of public opinion, as they offer a ``push-button magic solution'' to something that otherwise requires a high-investment skill, which can invalidate the time and effort that many professionals have spent learning their craft. The idea that traditional methods of mixing are still superior and that learning by trial and error is the best way to master the art of mixing is still quite prevalent among the expert community.

\begin{figure}
    \centering
    \includegraphics[scale=0.7]{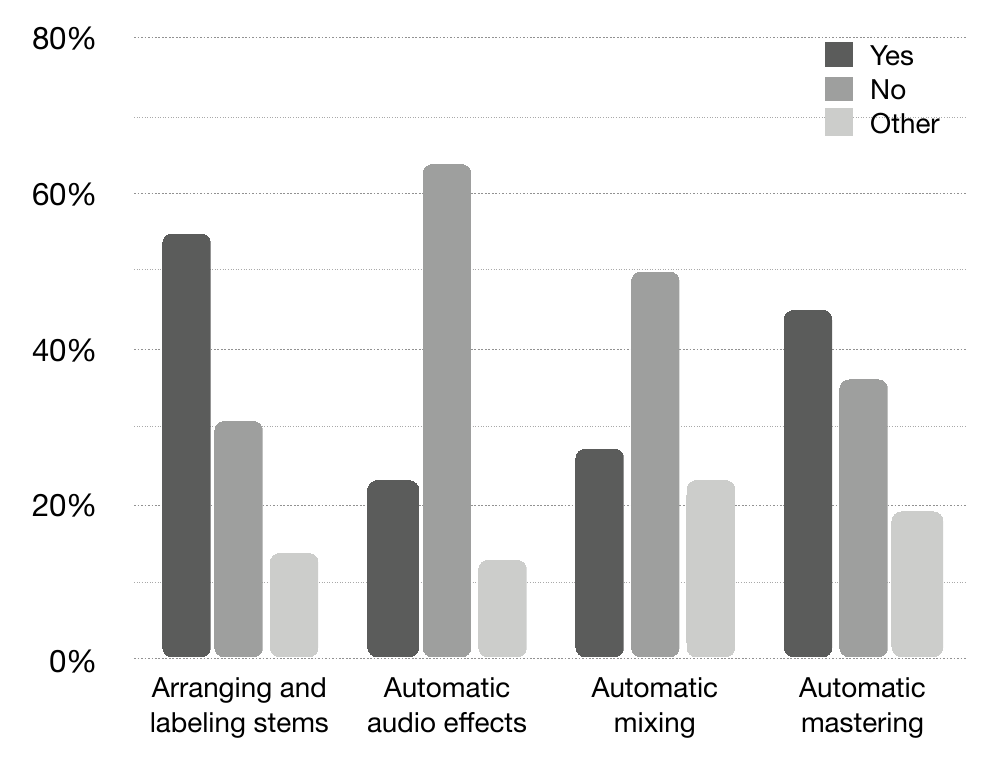}
    \caption{Preference for the use of different AI tools in mixing workflow as reported by pro-ams and pros}
    \label{fig:diffst}
\end{figure}

In conclusion, some professionals see the potential of AI in improving their workflows, but ultimately it is up to individual preferences and workflows. \Cref{fig:diffst} presents the preference for usage of different categories of AI tools in mixing workflows as reported by pro-ams and pros. We see that there is more acceptance for automatic mixing over tools that apply audio effects directly to the given recordings. This could be explained as most often these automatic audio effect tools do not understand the context of the given stem in the mix and are very generic in nature. We also observe that there is greater acceptance of labelling tools and automatic mastering tools. Many pros confirmed that they generally use auto-mastering tools to adjust the overall dynamics and loudness of a draft of the mix before sharing it with the client. They also explained that loading up multitrack in the DAW, labelling, and colour-coding stems before mixing is repetitive and takes up much time.

\section{Discussion}

\textit{ ``Mixing is without doubt an art, but there’s a lot that AI can help us with, and it’s an inevitable next step.''} [F]

Most amateurs want to share their music with the world. However, the bottleneck to successful commercial release most often is mixing. Honing the skills for mixing or hiring a professional engineer is often an expensive affair. This category of users might employ automatic mixing services to create mixes that they believe are good enough for them to target a commercial release. Amateurs might also find use in assistive mixing tools that are corrective and enable smart editing~\citep{Fazekas:2007hl}. This category of users are less interested in fine-grained control and customisation. 

Pro-ams, on the other hand, are more intentional and particular about the quality of the final result and may be producing and mixing music for themselves or others. Pro-ams encompass the entire spectrum of users between amateurs and professionals. As mixing is not their primary source of income, they are more exploratory and learning-focused. This category of users deploy AI-based tools in their workflows, however, given their higher levels of critical listening skills, they are looking for tools that are precise and offer more control. They use AI mixing tools as a starting point or for inspiration, but need more control over the final mix.

Finally, for professionals, mixing is an art form in itself. They are willing to spend more money on tools that they find useful and can help them speed up technical and repetitive tasks. However, they need a lot of control and customisation options. \Cref{Table:desired_tools} lists smart tools under various categories and tasks that professionals ascertained will ease their mixing workflows. It should also be noted that there is less enthusiasm for fully automated mixing tools amongst experts as compared to those that provide some control. This could be because automated mixing is often seen as a ``black box'' with limited control. Through our interviews, it became evident that most professionals are open to using assistive and collaborative technologies in their workflows.

\textit{``...but I think where the computer system is interacting with the human, that's where it's really interesting.''} [I]

We aggregate the use cases and expectations of various user groups from smart tools for mixing in \Cref{comparision}.


\begin{table*}[htb]
\centering
\renewcommand{\arraystretch}{1.5}
\begin{tabular}{|p{0.12\linewidth}|p{0.2\linewidth}|p{0.25\linewidth}|p{0.28\linewidth}|}
\hline
\textbf{Smart tools\textbackslash{}User} & \textbf{Amateurs}            & \textbf{Pro-ams}              & \textbf{Professionals}                           \\ \hline
\multirow{4}{*}{\textbf{Use-case}}       & Create decent mix            & Learning and exploratory tool & Automate repetitive and time-consuming tasks     \\ \cline{2-4} 
                   & As a learning tool & To find a starting point               & Co-creation and assistance                 \\ \cline{2-4} 
                   &                    & Automate technical tasks               & To find a starting point/direction for mix \\ \cline{2-4} 
                   &                    & Creativity and inspiration             & Creativity and inspiration                 \\ \hline
\multirow{5}{*}{\textbf{Expectations}}   & Autonomous with less control & Advanced and more control     & Highly advanced and wide range of control option \\ \cline{2-4} 
                   & Cost-effective     & Accurate and precise                   & Accurate and precise                       \\ \cline{2-4} 
                   & Easy to use        & Assistive                              & Assistive                                  \\ \cline{2-4} 
                   &                    & Cost effective                         & Easy integration in current workflow       \\ \cline{2-4} 
                   &                    & Easy integration into current workflow & Context-aware                              \\ \hline
\textbf{Sentiment} & Positive           & Cautiously positive                    & Mixed                                      \\ \hline
\end{tabular}
\caption{Comparison of use-case, expectations, and sentiment amongst different categories of users of AI technology in mixing workflows}
\label{comparision}
\end{table*}

\begin{table*}[htb]
\renewcommand{\arraystretch}{1.5}
\centering
\begin{tabular}{|p{0.15\linewidth}|p{0.25\linewidth}|p{0.5\linewidth}|}
\hline
\textbf{Phase} & \textbf{Category}                & \textbf{Tasks}                                          \\ \hline
\multirow{8}{*}{\textbf{Before Mixing}} & \multirow{4}{*}{Mix preparation*******} & Labelling and colour-coding tracks***   \\ \cline{3-3} 
               &                                  & Grouping and arranging tracks                           \\ \cline{3-3} 
               &                                  & Importing session data                                  \\ \cline{3-3} 
               &                                  & Setting up sends, AUX, and buses**                      \\ \cline{2-3} 
               & \multirow{4}{*}{Editing}         & Trimming Silences                                       \\ \cline{3-3} 
               &                                  & Identifying and fixing clicks and pops                  \\ \cline{3-3} 
               &                                  & Identifying and fixing phase issues               \\ \cline{3-3} 
               &                                  & Checking mono compatibility                             \\ \hline
\multirow{8}{*}{\textbf{During Mixing}} & \multirow{2}{*}{Levels Balancing}       & Gain staging**                          \\ \cline{3-3} 
               &                                  & Setting a rough fader mix***                            \\ \cline{2-3} 
                                        & \multirow{3}{*}{Spectral Corrections}   & Identifying and fixing masking issues** \\ \cline{3-3} 
               &                                  & Auto-high pass filter for content with low bass content \\ \cline{3-3} 
               &                                  & Auto EQ and EQ matching***                              \\ \cline{2-3} 
               & Communicating                    & Getting artist feedback on time-stamps                  \\ \cline{2-3} 
               & \multirow{2}{*}{Co-creativity**} & Suggesting ideas for mix**                              \\ \cline{3-3} 
               &                                  & Recommendations for audio effect processing             \\ \hline
\end{tabular}
\caption{List of smart functionalities that mix engineers desire to ease their workflow; Frequency of the requests for the tool is represented by the number of asterisks (*)}
\label{Table:desired_tools}
\end{table*}


\subsection{Future of smart technology for mixing}
The key takeaway of this investigation is that each user group has different goals and requirements, hence they might require tools that cater to their specific goals and needs. The future and utilisation of AI in mixing workflows are dependent on the design objectives of the tools. For the wider acceptance of these tools, it is important to build them such that they satisfy the three factors mentioned in \Cref{factors}. Here, we would like to propose a brief overview of possible directions.

\subsubsection{Balance of Control and Automation}
The right amount of control and automation has always been a matter of debate in human-computer interaction~(HCI)~\citep{xu2023transitioning, mcfarlane2002scope}. AI systems are often seen as black boxes that give out results that offer less control and interpretability. Our results show that each user group desires varying levels of automation and control. To counter this we propose designing AI tools that offer different levels of control and automation based on the user's level of expertise. This could be achieved by allowing the same plugin to have different modes, ranging from full automation, conditional automation, partial automation, assistive to no automation~\citep{tsiros2020towards}. This would allow the user to salvage the same plugin/tool for multiple purposes based on the time available and the precision required. There have been works in this direction to design AI-based mixing systems that are interpretable and offer more control to the end user~\citep{steinmetz2020learning}.

\subsubsection{Seamless Integration}
Professional mixing engineers often have an established workflow and familiar tools that helps them be efficient in their practice. Hence, we need to build tools and technology that have similar formats and configurations to what these users are familiar with. This implies turning research into products that are familiar, easy to access, navigate, and thus integrate into their existing workflows. For example, a model trained to apply automatic equalisation could be incorporated as a plugin that can be easily loaded into a DAW. This calls for the creation of developer tools that enable quicker and easier translation of research into technology. This can be achieved by using frameworks and software development kits~(SDK) like Neutone\footnote[6]{\url{https://neutone.space/}}, JUCE\footnote[7]{\url{https://juce.com/}}, CMajor\footnote[8]{\url{https://cmajor.dev/}}, iPlug2\footnote[9]{\url{https://iplug2.github.io/}}, and RTNeural\footnote[10]{\url{https://github.com/jatinchowdhury18/RTNeural}}. This would allow researchers to get quicker feedback from the creatives and thus iterate over to create better tools with a clearer vision of the user needs. 
Another potential approach could be where the AI learns to control a DAW. This implies that given the stems in a project session, the system is able to predict the gain levels, panorama values, and audio effect chains with their parameters for stems to create a mix. This could be helpful for hobbyists and low-budget musicians to improve their productions, while professionals would use it to save time and create a boilerplate mix that can be refined later. This also opens the door for remodelling and redesigning existing DAWs to be able to use the increased power of consumer laptops and embedded hardware available in professional mixing consoles~\citep{10061604}.

\subsubsection{Context-Aware Systems}
We observed that several pro-ams and pros admitted that AI-based tools are very generic and do not work well for outliers. They also expressed that these tools often only analyse the given track without understanding the context in the mix, which lowers trust and confidence in these tools.

\textit{"AI can't meaningfully interpret what your song calls for, its vibe. It can only do generic."} [F]

This opens up the possibility to develop systems that are context-aware~\citep{lefford2021context} and are trained on more diverse music and audio data to be able to work well for a variety of styles, genres and niche characteristics that are often a signature of innovation in the field. 

These AI systems could also be made more contextually aware of the vision of the mix by allowing them to be queried by text~\citep{agostinelli2023musiclm}, audio~\citep{koo2022music}, and semantic descriptors~\citep{stables2014safe, stables2016semantic,Turchet20}. Recent works have presented a state-of-the-art performance in text-based, melody-based, and semantic token-based music generation. We believe these systems could be further explored and fine-tuned for audio engineering applications.

Context-aware tools have the potential to bring in new talent, new genres and diversity into the music industry because of the ease of production that they offer. Amateurs might find it easier to produce and share music, thus potentially getting recognised for their craftsmanship. These tools might also help pro-ams improve skills and practice mixing professionally. Overall, these tools have the potential to improve mixing workflows and increase productivity and creativity.

\section{Summary}
The use of AI in mixing workflows is a complex and nuanced topic. While some see it as a potential game-changer, others are more skeptical about its ability to replace the human touch and creativity required in the mixing process. In this investigation, we identified three categories of users for the AI-based tool for mixing workflows. Based on the knowledge gained from them, we analysed and reported the sentiments and expectations of each of them. We argue that this information can be helpful and create a basis for further development and application of AI tools that can simplify the work of users in each category. 

In conclusion, we believe that the user-specific insight that our study provides will aid researchers in building AI mixing technologies that are better accepted and integrated by the audio engineering community in the time to come. Additionally, AI-powered tools could become further precise, versatile and adaptable to different genres and styles of music, and able to work with a wider range of input audio. Another possibility is the development of AI that can understand and respond to the creative intent of the user, allowing for more intuitive and efficient collaboration between humans and machines in the music production process. The field of AI and Music is constantly evolving, and the future possibilities are only limited by the current technology and the imagination of the researchers.

\section{Acknowledgements}
We express our sincere gratitude to the AES Europe convention reviewers for providing valuable feedback on our work. We are truly appreciative of the contributions made by all the mixing engineers and participants who participated in the interviews and studies and shared their valuable knowledge with us. We extend our thanks to the Steinberg research and development team for their unwavering support and honest feedback throughout this project.

Additionally, we acknowledge the invaluable input of our colleagues, Christian Steinmetz, Gary Bromham, and Angeliki Mourgela, who engaged in insightful discussions on this topic and provided critical feedback that helped shape our work.

We are also grateful for the financial support provided by UK Research and Innovation [grant number EP/S022694/1]. This funding has been instrumental in facilitating our research and enabling us to contribute to the field.

\bibliographystyle{jaes}

\bibliography{refs}

\end{document}